\documentclass[a4paper,floats]{jpconf}
\usepackage{graphicx}
\usepackage{amssymb}
\usepackage{dcolumn}
\usepackage{amsmath}
\usepackage{bm}
\usepackage{epsfig}
\usepackage{float}
\usepackage{caption}

\newcommand{\be}{\begin{equation}}
\newcommand{\ee}{\end{equation}}

\def\mod{ \mathop{\rm mod} }

\def\NN{\mathbb{N}}
\def\PN{\mathbb{E}}
\def\QN{\mathbb{O}}

\begin{document}
\title{Criticality and Chaos in Systems of Communities}

\author{Massimo Ostilli and Wagner Figueiredo}

\address{
Departamento de Fisica, Universidade Federal de Santa Catarina, Florianopolis, 88040-900, SC, Brazil}

\ead{massimo.ostilli@gmail.com}

\begin{abstract}
We consider a simple model of communities interacting via bilinear terms. 
After analyzing the thermal equilibrium case, which can be described by an Hamiltonian,
we introduce the dynamics that, for Ising-like variables, reduces to a Glauber-like dynamics.
We analyze and compare four different versions of the dynamics: flow (differential equations), map (discrete-time dynamics), 
local-time update flow, and local-time update map. The presence of only bilinear interactions prevent
the flow cases to develop any dynamical instability, the system converging always to the thermal equilibrium. 
The situation is different for the map when unfriendly couplings are involved,
where period-two oscillations arise. In the case of the map with local-time updates, oscillations of any period and chaos can
arise as a consequence of the reciprocal ``tension'' accumulated among the communities during their sleeping time interval.
The resulting chaos can be of two kinds: true chaos characterized by positive Lyapunov exponent and bifurcation cascades, 
or marginal chaos characterized by zero Lyapunov exponent and critical continuous regions.
\end{abstract}

\section{Introduction}
The fact that the speed of physical interactions is finite, as for light or sound, has well known crucial consequences in all physics, 
and it is the ultimate reason for the existence of waves. For example, if we artificially assume the speed of light as infinite 
in the Maxwell equations, the magnetic field would be zero and we would not have electromagnetic waves.

The situation is quite different when the equations contain non linear terms,
which can result in erratic or even chaotic trajectories.
Notice that, in the absence of mechanical kinetic terms, which is our framework, the presence of non linear terms
is exclusively due to the presence of non bilinear interactions among the components of the system. 
For example, in the Kuramoto model \cite{KuramotoC}-\cite{KuramotoR}, 
sinusoidal interactions lead to chaos \cite{StrogatzC}-\cite{Volatility} if the coupling constant $K$ is larger than some threshold $K_c$.

It should be emphasized that the presence of non linear terms is not always sufficient for the emergence of chaos.
However, it has been observed that the introduction of time-delays among the components of a system
augments the chance that the system develops chaos \cite{Glass}-\cite{Sprott}. 
The idea at the base of such a mechanism is that time-delays among the system components makes it difficult for the
system to set in a steady state. 
In particular, for linear flows, a fixed delay among the system components corresponds to a finite speed of interactions,
which in turn leads to oscillations or waves. 
For non linear flows a delay can have more dramatic consequences, with chaos scenarios more probable.

In \cite{ChaosMO}, we have introduced a simple discrete-time two components dynamical model, characterized
by local-time updates. Local time updates means that the each component updates its status only at certain times $t$.
In the specific model analyzed in \cite{ChaosMO}, a component updates (sleeps) its status only at odd times $t=1,3,\ldots$, 
(even times $t=0,2,\ldots$), while the other component updates (sleeps) 
its status only at even times $t=0,2,\ldots$ (odd times $t=1,3,\ldots$). 
Remarkably, as we have shown in \cite{ChaosMO}, for certain coupling values, this simple system     
exhibits chaos even if it is characterized only by bilinear interactions. 
In fact, when the couplings are in the steady state regime, the system can be described via a Hamiltonian which
is quadratic in the system components.
We are not aware of a similar case in the literature. 
Our understanding here is that there are three factors generates chaos: 
\textit{i)} the presence of unfriendly couplings, \textit{ii)} the discrete-time nature of the dynamics,
and \textit{iii)} the existence of at least two interacting communities which rearrange their configurations
at alternate times, \textit{i.e.} via local-time updates.
Notice that \textit{(iii)} is not just a delay; it also implies that each component has a sleeping period.
It is during such a sleeping period that the component accumulates a tension with respect to the other component when
the reciprocal coupling is unfriendly.
We are in fact dealing with a sort of dynamical frustration.   

In this paper we review some aspects of the model which were only partially treated in \cite{ChaosMO}.
In particular, we compare the analogous flow model with the present map and show that the above
conditions \textit{(i)}-\textit{(iii)} are in fact necessary, and calculate the maximal Lyapunov exponent.

The reader not interested in the derivation of the dynamic equations may skip the next Section.

\section{The model in the flow, map and local-time versions}
Here we consider a two components system. Generalization to an arbitrary number of components will be reported elsewhere.
We consider two communities \cite{Santo} of agents $\mathcal{N}^{(1)}$ and $\mathcal{N}^{(2)}$,  
with cardinalities $N^{(1)}$ and $N^{(2)}$.
The agent with index $i$ can be in two possible status, $\sigma_i=\pm 1$, with $i\in \mathcal{N}^{(1)}\cup \mathcal{N}^{(2)}$. 
According to the sign of the couplings, friendly or unfriendly, 
each agent $i$ tends to follow or anti-follow its neighbors, by minimizing or maximizing the term
$\sigma_i\sum'\sigma_j$, where $\sum'$ runs over the set of neighbors of $i$.
It is necessary to distinguish between intra- and inter-couplings,
therefore we introduce the $2\times 2$ matrix $J^{(l,m)}$: 
$J^{(1,1)}$ and $J^{(2,2)}$ are the intra-couplings, and $J^{(1,2)}=J^{(2,1)}$ the inter-coupling. 
%for the two intra-couplings $J^{(1,1)}$ and $J^{(2,2)}$, and the the inter-coupling $J^{(1,2)}=J^{(2,1)}$. 
In the most general formulation
we should introduce also the $2\times 2$ matrix $\Gamma^{(l,m)}$
defined as the set of coupled spins $(i,j)$ within the same community (intra), or between the two communities (inter).
Finally, we introduce a global factor $\beta$ that rescales all the couplings in $\beta J^{(l,m)}$.

Let us indicate the time by $t$, which is supposed to range in the set of natural numbers $\NN$ for the map, while
it can take any non negative real number for the flow.
Inspired by the Glauber dynamics \cite{Glauber}, 
for flow and map we introduce the following transition rate probabilities for the spin with state $\sigma_i$ to jump to the state~$\sigma_i'$
\begin{eqnarray}
\label{rate0}
w^{(\mathrm{FLOW/MAP})}(\sigma_i\to\sigma_i';t)=   \left\{ 
%\nonumber
\begin{array}{l}
\frac{1+\sigma_i'\tanh\left(\beta J^{(1,1)}\sum_{j:(i,j)\in \Gamma^{(1,1)}}\sigma_j+
\beta J^{(1,2)}\sum_{j:(i,j)\in \Gamma^{(1,2)}}\sigma_j\right)}{2}, ~\mathrm{for~} i\in\mathcal{N}^{(1)},\\
\frac{1+\sigma_i'\tanh\left(\beta J^{(2,2)}\sum_{j:(i,j)\in \Gamma^{(2,2)}}\sigma_j+
\beta J^{(2,1)}\sum_{j:(i,j)\in \Gamma^{(2,1)}}\sigma_j\right)}{2}, ~\mathrm{for~} i\in\mathcal{N}^{(2)}.
\end{array}
\right. 
\end{eqnarray} 

Let $(\PN,\QN)$ be any periodic partition of $\NN$, \textit{i.e.}, $\PN\cup \QN=\NN$, $\PN\cap\QN=\emptyset$, and 
$\PN$ and $\QN$ contain both infinite elements of $\NN$. 
For instance, $\PN$ and $\QN$ can be the set of even and odd numbers, respectively,
and we will assume this choice in all the next examples.
We now introduce the following local-time transition rate probabilities for the spin with state $\sigma_i$ to jump to the state~$\sigma_i'$

\begin{eqnarray}
\label{rate}
%w(\sigma_i\to\sigma_i';t)=~~~~~~~~~~~~~~~~~~~~~~~~~~~~~~~~~~~~~~~~~~~~~~~~~~~~~~~~ \nonumber \\  
w^{(\mathrm{LOCAL-TIME~FLOW})}(\sigma_i\to\sigma_i';t)=   
\left\{ 
%\nonumber
\begin{array}{l}
\delta_{\sigma_i,\sigma_i'} 
\quad i\in\mathcal{N}^{(1)}, \quad \mathrm{if~}\mod[t,2]\neq 0, \\ \\
\frac{1+\sigma_i'\tanh\left(\beta J^{(1,1)}\sum_{j:(i,j)\in \Gamma^{(1,1)}}\sigma_j+
\beta J^{(1,2)}\sum_{j:(i,j)\in \Gamma^{(1,2)}}\sigma_j\right)}{2}, 
\quad \\ \mathrm{for~} i\in\mathcal{N}^{(1)}, \quad \mathrm{if~}\mod[t,2]=0,\\ \\
\delta_{\sigma_i,\sigma_i'}
\quad i\in\mathcal{N}^{(2)}, \quad \mathrm{if~}\mod[t,2]\neq 0,\\ \\
\frac{1+\sigma_i'\tanh\left(\beta J^{(2,2)}\sum_{j:(i,j)\in \Gamma^{(2,2)}}\sigma_j+
\beta J^{(2,1)}\sum_{j:(i,j)\in \Gamma^{(2,1)}}\sigma_j\right)}{2}, 
\quad \\ \mathrm{for~} i\in\mathcal{N}^{(2)}, \quad \mathrm{if~}\mod[t,2]=0,
\end{array}
\right. 
\end{eqnarray} 
%Similarly, for the map we introduce
\begin{eqnarray}
\label{rate1}
%w(\sigma_i\to\sigma_i';t)=~~~~~~~~~~~~~~~~~~~~~~~~~~~~~~~~~~~~~~~~~~~~~~~~~~~~~~~~ \nonumber \\  
w^{(\mathrm{LOCAL-TIME~MAP})}(\sigma_i\to\sigma_i';t)=   
\left\{ 
%\nonumber
\begin{array}{l}
\delta_{\sigma_i,\sigma_i'} 
\quad i\in\mathcal{N}^{(1)}, \quad \mathrm{for~}t\in \QN, \\ \\
\frac{1+\sigma_i'\tanh\left(\beta J^{(1,1)}\sum_{j:(i,j)\in \Gamma^{(1,1)}}\sigma_j+
\beta J^{(1,2)}\sum_{j:(i,j)\in \Gamma^{(1,2)}}\sigma_j\right)}{2}, 
\quad \\ \mathrm{for~} i\in\mathcal{N}^{(1)}, \quad t\in \PN,\\ \\
\delta_{\sigma_i,\sigma_i'}
\quad i\in\mathcal{N}^{(2)}, \quad \mathrm{for~} t\in \PN,\\ \\
\frac{1+\sigma_i'\tanh\left(\beta J^{(2,2)}\sum_{j:(i,j)\in \Gamma^{(2,2)}}\sigma_j+
\beta J^{(2,1)}\sum_{j:(i,j)\in \Gamma^{(2,1)}}\sigma_j\right)}{2}, 
\quad \\ \mathrm{for~} i\in\mathcal{N}^{(2)}, \quad t\in \QN.
\end{array}
\right. 
\end{eqnarray} 

From the viewpoint of modeling, Eqs. (\ref{rate0}), (\ref{rate}), or (\ref{rate1}), are justified as 
they make each spin to follow the majority of its intra- and inter-neighbors and,
thanks to the presence of the functions $\tanh(\cdot)$, the rates are non negative
and normalized at any time $\sum_{\sigma'}w(\sigma\to\sigma';t)=1/$(Time Unit).
From a deeper viewpoint, Eqs. (\ref{rate0}), (\ref{rate}), or (\ref{rate1}),
are based on the fact that, as we will see soon, in the case of positive couplings 
they lead to Boltzmann equilibrium governed by only quadratic interactions.
More precisely, in the case of positive couplings
these forms guarantee that at equilibrium the system satisfies 
the principle of detailed balance and the principle of
maximal entropy for any quadratic interactions.

We formalize the discrete-time probabilistic dynamics induced by Eqs. (\ref{rate0}), (\ref{rate}), or (\ref{rate1}), or  as follows. 
Let $N=N^{(1)}+N^{(2)}$. Let us introduce the 
spin vector $\bm{\sigma}=(\sigma_1,\ldots,\sigma_N)$, and the 
associated probability vector $p(\bm{\sigma};t)$,
\textit{i.e.}, the probability that the system
is in the configuration $\bm{\sigma}$ at time $t\in \NN$.
The master equation for flow and map (local-time or not) reads
%\begin{flalign}
\begin{eqnarray}
\label{Master1}
&& \frac{\partial p^{(\mathrm{FLOW})}(\bm{\sigma};t)}{\partial (\alpha t)}=
-\sum_{\bm{\sigma}^{'}}p^{(\mathrm{FLOW})}(\bm{\sigma};t)W(\bm{\sigma}\to\bm{\sigma}^{'}) %~~~~ \nonumber \\ && 
%\\ && 
+\sum_{\bm{\sigma}^{'}}p^{(\mathrm{FLOW})}(\bm{\sigma}^{'};t)W(\bm{\sigma}^{'}\to\bm{\sigma}),
\end{eqnarray}
\begin{eqnarray}
\label{Master1}
&& \frac{p^{(\mathrm{MAP})}(\bm{\sigma};t+1)-p^{(\mathrm{MAP})}(\bm{\sigma};t)}{\alpha}=
-\sum_{\bm{\sigma}^{'}}p^{(\mathrm{MAP})}(\bm{\sigma};t)W(\bm{\sigma}\to\bm{\sigma}^{'}) \nonumber \\ && 
+\sum_{\bm{\sigma}^{'}}p^{(\mathrm{MAP})}(\bm{\sigma}^{'};t)W(\bm{\sigma}^{'}\to\bm{\sigma}),
\end{eqnarray}
%\end{flalign}  
where we have introduced the global transition rates% in terms of the local ones
\begin{eqnarray}
\label{W}
W(\bm{\sigma}\to\bm{\sigma}^{'})=\prod_{i\in \mathcal{N}^{(1)}\cup \mathcal{N}^{(2)}} w(\sigma_i\to\sigma_i'),
\end{eqnarray}
where $w(\sigma_i\to\sigma_i')$ can be of the type Eqs. (\ref{rate0}), (\ref{rate}), or (\ref{rate1}),
and where $\alpha/2>0$ may be interpreted as the rate at which,
due to the interaction with an environment,
a free spin ($J^{(l,m)}=0$) makes transitions from either state to
the other. As we have proved in \cite{DPotts}, it is necessary
to impose the bound $\alpha\leq 1$ for $p(\bm{\sigma};t)$ to 
be a probability at any time $t$.
By using Eqs. (\ref{rate})-(\ref{W}) it is easy to check that the stationary solutions
$p(\bm{\sigma})$ of Eq. (\ref{Master1}) are given by the Boltzmann distribution 
$p(\bm{\sigma})\propto \exp[-\beta H (\bm{\sigma})]$, where 
\begin{eqnarray}
\label{H}
H= && - J^{(1,1)}\sum_{(i,j)\in\Gamma^{(1,1)}}\sigma_i\sigma_j-J^{(2,2)}\sum_{(i,j)\in\Gamma^{(2,2)}}\sigma_i\sigma_j%\nonumber \\ &&
      - J^{(1,2)}\sum_{(i,j)\in\Gamma^{(1,2)}}\sigma_i\sigma_j.
\end{eqnarray}
Of course, the existence of a stationary solution $p(\bm{\sigma})$
does not represent a sufficient condition for equilibrium.
In fact, 
asymptotically the system can reach non-point-like attractors, and even aperiodic or chaotic regimes.
However, as we have discussed in the Introduction, the fact that for certain values of couplings
the system can be described via the Hamiltonian (\ref{H}), means that we are dealing
with only bilinear interactions. 
Here, the non linearities that lead to chaos are caused by the local time updates.

From now on, we will omit the suffixes (FLOW), (MAP) and (LOCAL-TIME). It will be clear from
the context what we are referring to.
Eqs. (\ref{rate})-(\ref{W}) define the microscopic dynamics from which
one can derive the macroscopic 
dynamics, \textit{i.e.}, the dynamics for the order parameters
%Eqs. (\ref{rate})-(\ref{W}) define the microscopic dynamics.
%From the microscopic dynamics one can derive the macroscopic (or reduced)
%dynamics, \textit{i.e.}, the dynamics followed by the order parameters
\begin{eqnarray}
\label{mm1}
&& x^{(1)}(t)=\sum_{\bm{\sigma}}p(\bm{\sigma};t)\frac{1}{N}\sum_{i\in\mathcal{N}^{(1)}} \sigma_i, \\
\label{mm2}
&& x^{(2)}(t)=\sum_{\bm{\sigma}}p(\bm{\sigma};t)\frac{1}{N}\sum_{i\in\mathcal{N}^{(2)}} \sigma_i.
\end{eqnarray}

\section{Mean Field Limit}
The mean-field limit is defined by the settings
$|\Gamma^{(1,1)}|=\binom{N^{(1)}}{2}$, $|\Gamma^{(2,2)}|=\binom{N^{(2)}}{2}$,
$|\Gamma^{(1,2)}|=N^{(1)}N^{(2)}$, and the replacements
$J^{(1,1)}\to J^{(1,1)}/N^{(1)}$, $J^{(2,2)}\to J^{(2,2)}/N^{(2)}$, $J^{(1,2)}\to J^{(1,2)}(N^{(1)}+N^{(2)})/(2N^{(1)}N^{(2)})$.
We parametrize the size of the two communities as 
\begin{eqnarray}
\label{aa}
N^{(1)}=N\rho^{(1)}, \quad N^{(2)}=N\rho^{(2)}, \quad \rho^{(1)}+\rho^{(2)}=1.
\end{eqnarray}  
Let us introduce the matrix $\bm{\tilde{J}}$ 
\begin{eqnarray}
\label{Jtilde}
\bm{\tilde{J}}=\left(
\begin{array}{ll}
 J^{(1,1)} & \frac{J^{(1,2)}}{2\rho^{(1)}} \\
\frac{J^{(2,1)}}{2\rho^{(2)}} & J^{(2,2)}
\end{array}
\right).
\end{eqnarray}  
As shown in \cite{ChaosMO}, for $N\to\infty$ 
%\end{widetext}
we obtain
the following deterministic evolution Eqs. for the order parameters (\ref{mm1})-(\ref{mm2})
\subsection{FLOW}
\begin{eqnarray}
\label{FF1}
&&\frac{\partial x^{(1)}(t)}{\partial (\alpha t)}=
\tanh\left(\beta \tilde{J}^{(1,1)}x^{(1)}(t)+\beta \tilde{J}^{(1,2)}x^{(2)}(t)\right)-x^{(1)}(t),
\end{eqnarray}  
\begin{eqnarray}
\label{FF2}
&&\frac{\partial x^{(2)}(t)}{\partial (\alpha t)}=
\tanh\left(\beta \tilde{J}^{(2,2)}x^{(2)}(t)+\beta \tilde{J}^{(2,1)}x^{(1)}(t)\right)-x^{(2)}(t),
\end{eqnarray}  

\subsection{MAP}
\begin{eqnarray}
\label{MM1}
&&\frac{x^{(1)}(t+1)-x^{(1)}(t)}{\alpha}=
\tanh\left(\beta \tilde{J}^{(1,1)}x^{(1)}(t)+\beta \tilde{J}^{(1,2)}x^{(2)}(t)\right)-x^{(1)}(t),
\end{eqnarray}  
\begin{eqnarray}
\label{MM2}
&&\frac{x^{(2)}(t+1)-x^{(2)}(t)}{\alpha}=
\tanh\left(\beta \tilde{J}^{(2,2)}x^{(2)}(t)+\beta \tilde{J}^{(2,1)}x^{(1)}(t)\right)-x^{(2)}(t),
\end{eqnarray}  

\subsection{LOCAL-TIME FLOW}
\begin{eqnarray}
\label{FFL1}
&&\frac{\partial x^{(1)}(t)}{\partial (\alpha t)}=%\\ &&
\left\{
\begin{array}{lll}
0,  \quad \mathrm{if~}\mod[t,2]\neq 0, \\ \\
\tanh\left(\beta \tilde{J}^{(1,1)}x^{(1)}(t)+\beta \tilde{J}^{(1,2)}x^{(2)}(t)\right)-x^{(1)}(t),~\mathrm{if~}\mod[t,2]=0,
\end{array}
\right.
\end{eqnarray}  
\begin{eqnarray}
\label{FFL2}
&&\frac{\partial x^{(2)}(t)}{\partial (\alpha t)}=%\\ &&
\left\{
\begin{array}{lll}
0,  \quad \mathrm{if~}\mod[t,2]\neq 0,\\ \\
\tanh\left(\beta \tilde{J}^{(2,2)}x^{(2)}(t)+\beta \tilde{J}^{(2,1)}x^{(1)}(t)\right)-x^{(2)}(t),~\mathrm{if~}\mod[t,2]=0.
\end{array}
\right.
\end{eqnarray}  

\subsection{LOCAL-TIME MAP}
\begin{eqnarray}
\label{MML1}
&&\frac{x^{(1)}(t+1)-x^{(1)}(t)}{\alpha}=%\\ &&
\left\{
\begin{array}{lll}
0,  \quad t\in \QN, \\ \\
\tanh\left(\beta \tilde{J}^{(1,1)}x^{(1)}(t)+\beta \tilde{J}^{(1,2)}x^{(2)}(t)\right)-x^{(1)}(t),~t\in \PN,
\end{array}
\right.
\end{eqnarray}  
\begin{eqnarray}
\label{MML2}
&&\frac{x^{(2)}(t+1)-x^{(2)}(t)}{\alpha}=%\\ &&
\left\{
\begin{array}{lll}
0,  \quad t\in \PN,\\ \\
\tanh\left(\beta \tilde{J}^{(2,2)}x^{(2)}(t)+\beta \tilde{J}^{(2,1)}x^{(1)}(t)\right)-x^{(2)}(t),~t\in \QN.
\end{array}
\right.
\end{eqnarray}

\subsection{Stationary solutions}
When the couplings are positive,
there is little difference among the four dynamics,
and they all asymptotically reach 
equilibrium according to the Boltzmann distribution  
$p(\bm{\sigma})\propto \exp[-\beta H_{\mathrm{mf}} (\bm{\sigma})]$, where 
$H_{\mathrm{mf}} (\bm{\sigma})]$ is the mean-field analogous of Eq. (\ref{H}),
%\begin{eqnarray}
%\label{Hmf}
%&& H_{\mathrm{mf}}=-J^{(1,1)}\sum_{(i,j)\in\Gamma^{(1,1)}}\sigma_i\sigma_j-J^{(2,2)}\sum_{(i,j)\in\Gamma^{(2,2)}}\sigma_i\sigma_j\nonumber \\
%&& -J^{(1,2)}\sum_{(i,j)\in\Gamma^{(1,2)}}\sigma_i\sigma_j,
%\end{eqnarray}
and $x^{(1)}(t)$ and $x^{(2)}(t)$ tend, for $t\to\infty$, to the stationary solutions of Eqs. (\ref{MM1})-(\ref{FFL2}), \textit{i.e.},
\begin{eqnarray}
\label{mstat}
\left\{
\begin{array}{lll}
x^{(1)}=\tanh\left(\beta \tilde{J}^{(1,1)}x^{(1)}+\beta \tilde{J}^{(1,2)}x^{(2)}\right),\\
x^{(2)}=\tanh\left(\beta \tilde{J}^{(2,1)}x^{(1)}+\beta \tilde{J}^{(2,2)}x^{(2)}\right).
\end{array}
\right.
\end{eqnarray}  

Eqs. (\ref{mstat}) represent a particular case of the general result derived in \cite{Comm}
valid for $n$ interacting communities at equilibrium.
In particular, one can check that Eqs. (\ref{mstat}) give rise to  
second order phase transitions whose critical surface is determined by the condition
\begin{eqnarray}
\label{crit}
\det\left(\bm{1}-\beta\bm{\tilde{J}}\right)=0.
\end{eqnarray}  
In general, $(x^{(1)},x^{(2)})=(0,0)$ is stable when the eigenvalues of $\beta \bm{\tilde{J}}$ 
are inside the interval $(-1,1)$ (disordered phase, or no consensus),
otherwise the system reaches a spontaneous ordered phase $(x^{(1)},x^{(2)})\neq (0,0)$ (frozen phase, or consensus). 
However, Eq. (\ref{crit}) is exact only for the flow and the map. In general, due to the sleeping time, 
the critical value of $\beta$ can differ from the one given by Eq. (\ref{crit}). This was erroneously not stressed in \cite{ChaosMO}.
The correct way to to derive the exact critical value of $\beta$ must use the full expression of the Jacobian, which is
the subject of Section 5. 

\section{Comparing flow, map and local-time versions}
From now on we shall always assume $\alpha=1$ and $N^{(1)}=N^{(2)}$.

In this Section we compare the four dynamics numerically. 

In Figs (1a)-(2b) we compare the normal flow and map. 
In Figs. (1a) and (1b) we consider a situation with
positive couplings at small and large values of $\beta$;
in Figs. (2a) and (2b) we consider a situation with
positive and negative couplings at small and large values of $\beta$, respectively.
It is evident that, if the couplings are positive, the flow and the map remain roughly close at any time,
and tend to coincide for $t\to\infty$ to the same stationary solution provided by Eq. (\ref{mstat}).
When instead some of the couplings are negative, at small $\beta$ the map tends to make damping oscillations around 
the flow (which for the present case is null), whereas, at large $\beta$, the map 
tends to make regular oscillations around the
flow solution (which for the present case is null).
\begin{figure}[htb]
\begin{center}
{\includegraphics[height=2.8in]{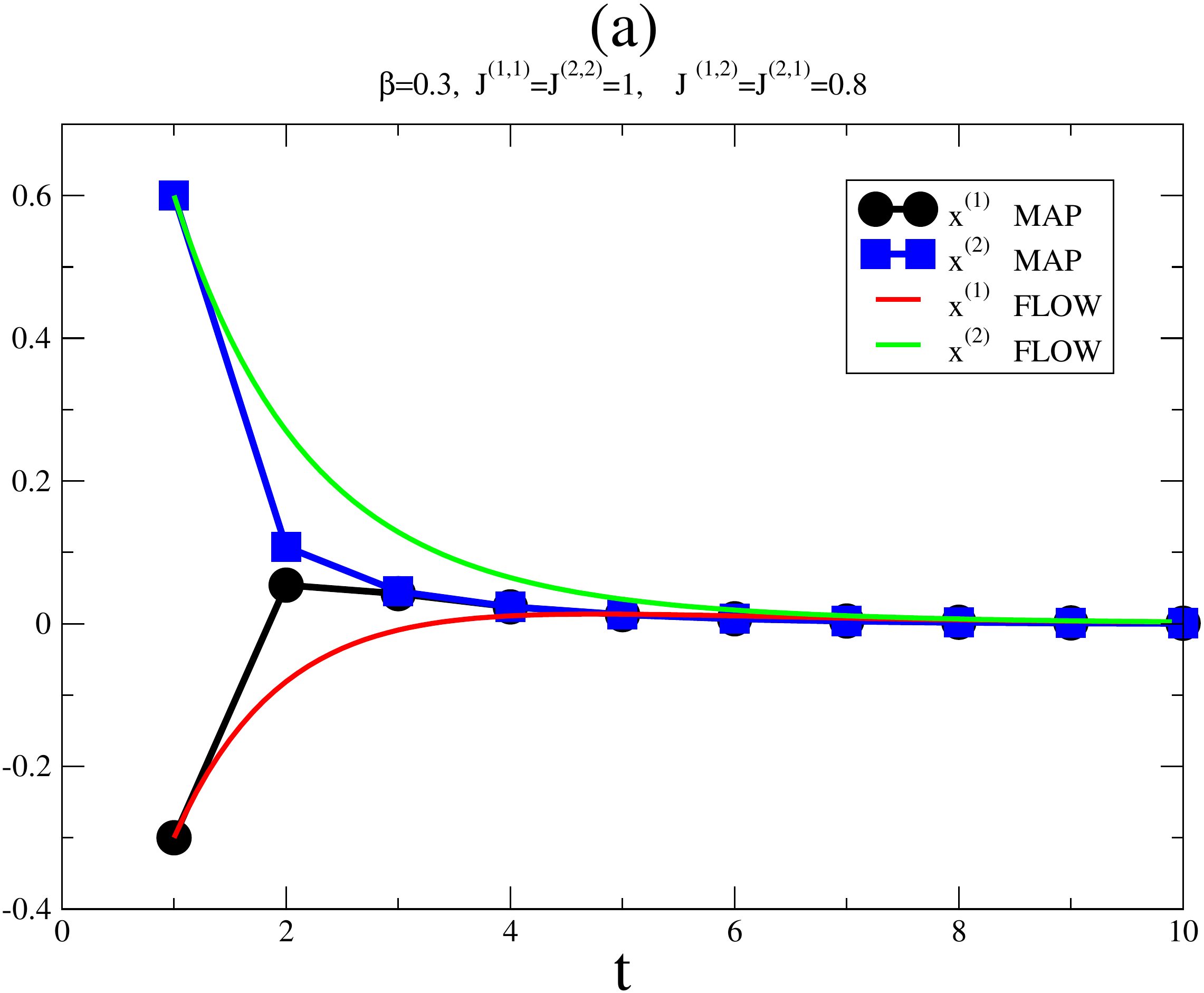}}\\  \vspace{0.5cm}
{\includegraphics[height=2.8in]{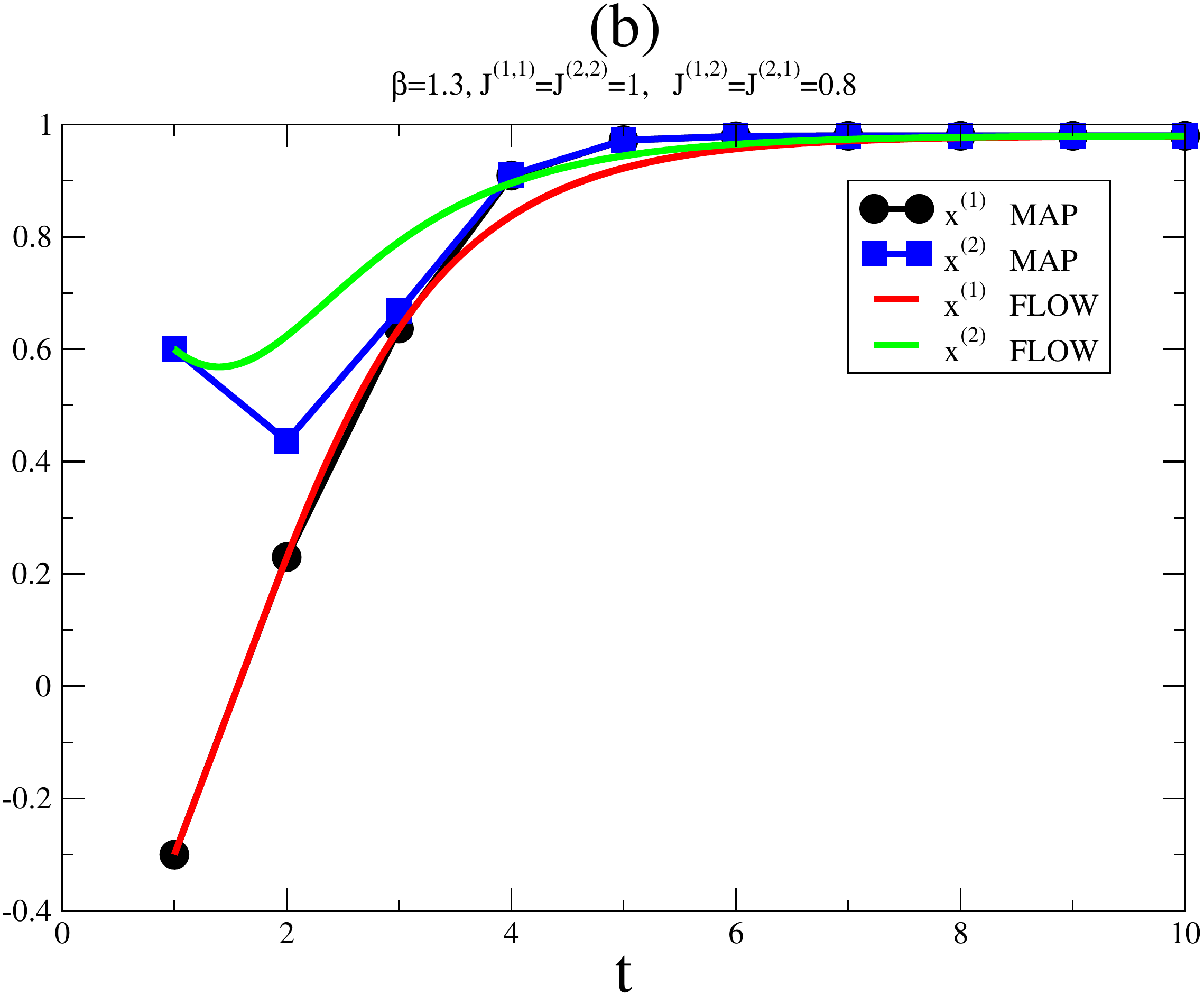}}
    \caption{(Color online) Plot of $x^{(1)}(t)$ and $x^{(2)}(t)$, solution of the flow, Eqs. (\ref{FF1})-(\ref{FF2}), 
and the map, Eqs. (\ref{MM1})-(\ref{MM2}) (here with $\alpha=1$), 
for positive couplings; below (Panel a) and above (Panel b) the threshold $\beta_c$.
Initial conditions are the same in all cases. 
The lines of the map are guides for the eyes. 
  \label{fig1}
    }
  \end{center}
\end{figure}

\begin{figure}[htb]
\begin{center}
{\includegraphics[height=2.8in]{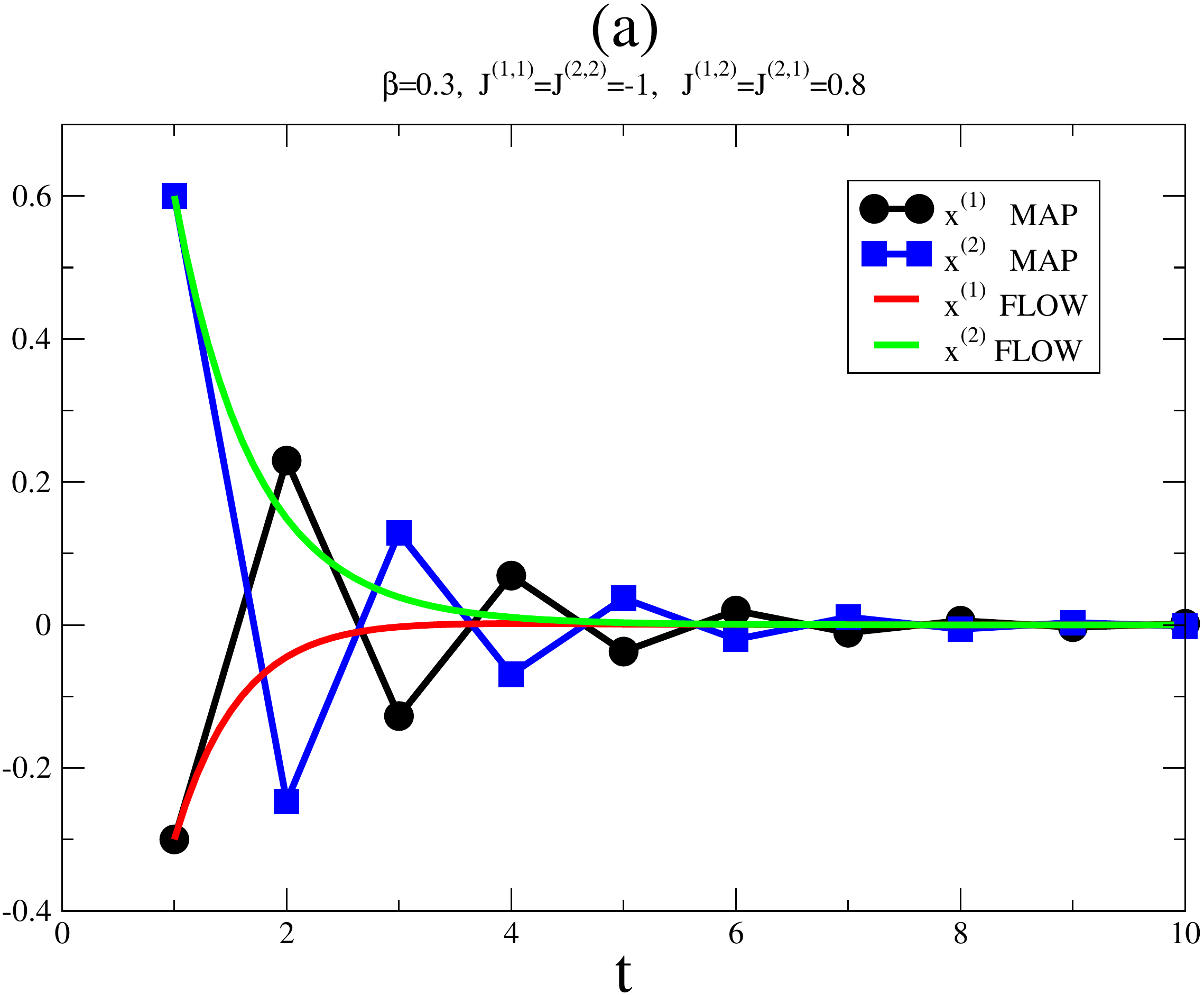}}\\ \vspace{0.5cm}
{\includegraphics[height=2.8in]{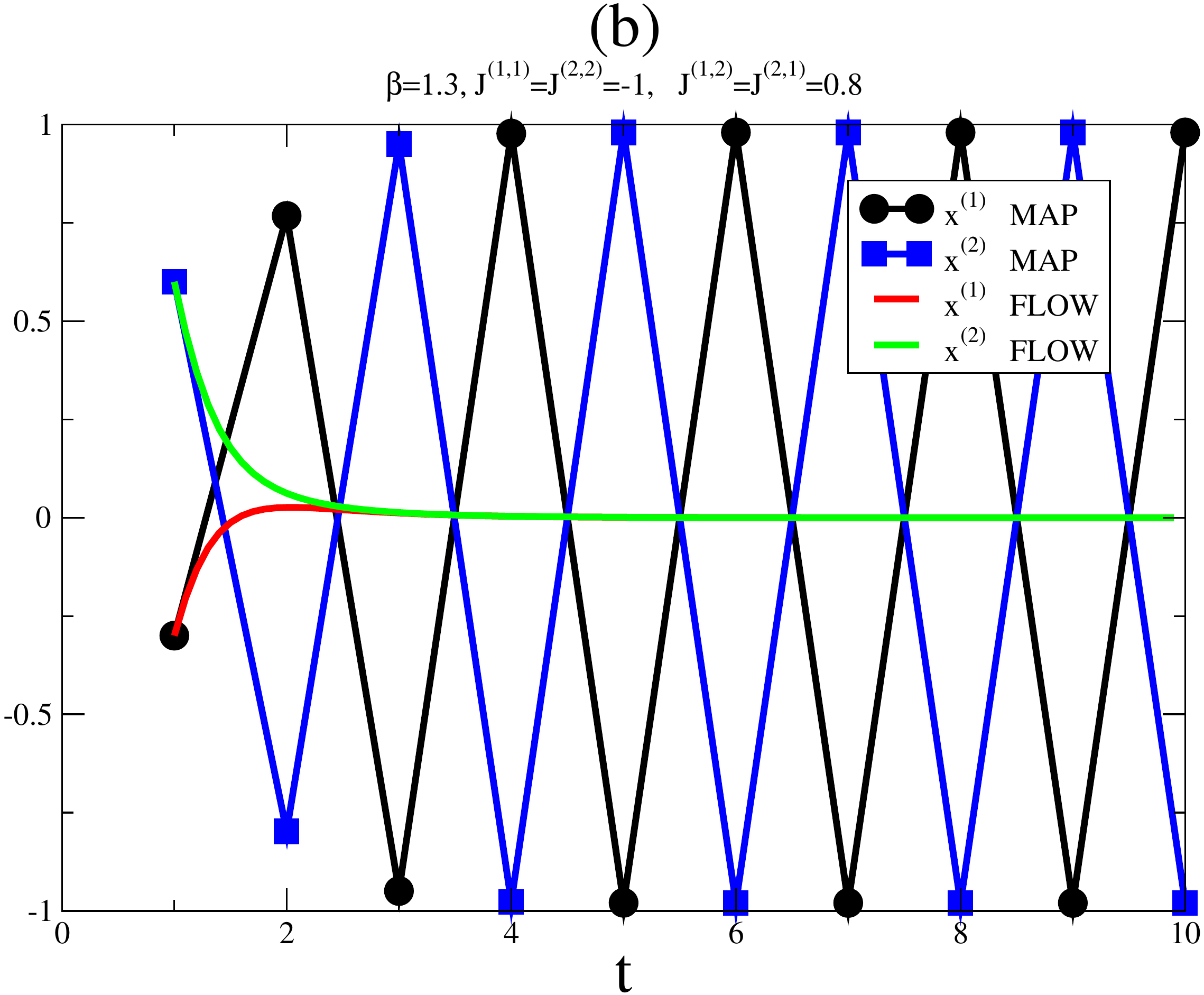}}
    \caption{(Color online) Plot of $x^{(1)}(t)$ and $x^{(2)}(t)$, solution of the flow, Eqs. (\ref{FF1})-(\ref{FF2}), 
and the map, Eqs. (\ref{MM1})-(\ref{MM2}) (here with $\alpha=1$),   
for a case with positive and negative couplings; below (Panel a) and above (Panel b) the threshold $\beta_c$.
Initial conditions are the same in all cases. 
The lines of the map are guides for the eyes. 
  \label{fig2}
    }
  \end{center}
\end{figure}

In Figs. (3a)-(3c) we compare the flow and map with local-time updates for 
a case with positive and negative couplings and three values of $\beta$, respectively.
Fig. (3a) provides the behavior for $\beta=1.5$ where the map follows damping oscillations
and both the map and the flow tend to the same stationary value provided by Eq. (\ref{mstat}).
Fig. (3b) provides the behavior for a relatively larger value of $\beta$, $\beta=2.5$, 
where the map follows regular oscillations of period 2, while the flow tends to a stationary value.
Fig. (3c) provides the behavior for $\beta=4.5$, 
where the map follows regular oscillations of period 4, while the flow tends to a stationary value.
Increasing further $\beta$ leads to larger and larger periods (not shown).

%\begin{figure}[htb]
%\begin{center}
%{\includegraphics[height=2.8in]{q_2_local_time.pdf}}\\ \vspace{0.5cm}
%{\includegraphics[height=2.8in]{q_2_local_time_period_2.pdf}}\\ \vspace{0.5cm}
%{\includegraphics[height=2.8in]{q_2_local_time_period_4.pdf}}
%    \caption{(Color online) Plot of $x^{(1)}(t)$ and $x^{(2)}(t)$, solution of the local-time flow, Eqs. (\ref{FFL1})-(\ref{FFL2}), 
%and the local-time map, Eqs. (\ref{MML1})-(\ref{MML2}) (here with $\alpha=1$),  
%$\beta=1.5$ (a), $\beta=2.5$ (b), and $\beta=4.5$ (c), all with the same initial conditions.
%Initial conditions are the same in all cases. 
%The lines of the map are guides for the eyes. 
%  \label{fig3}
%    }
%  \end{center}
%\end{figure}

\noindent%
\begin{minipage}{\linewidth}% to keep image and caption on one page
\begin{center}
{\includegraphics[height=2.8in]{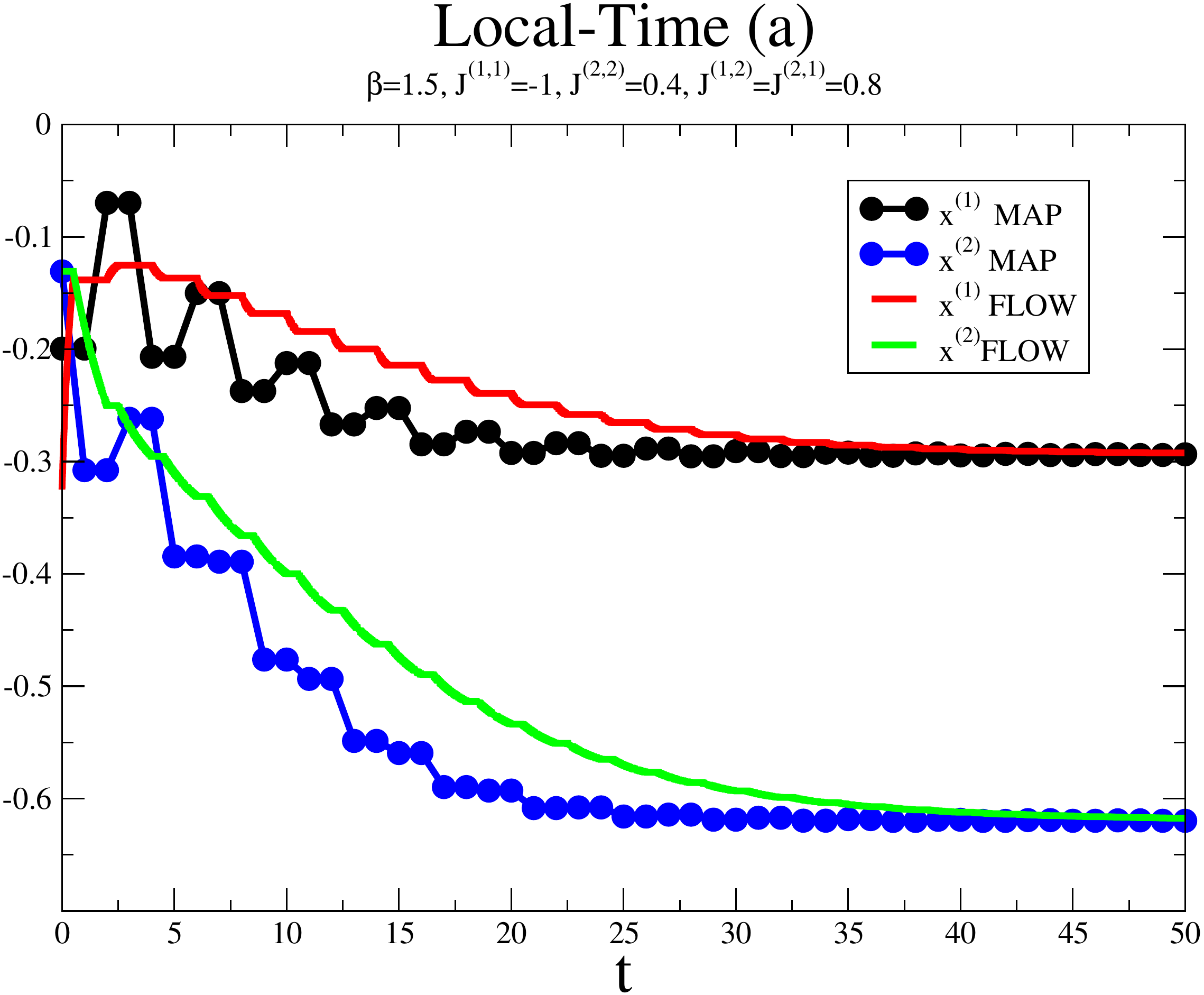}}\\ \vspace{0.5cm}
{\includegraphics[height=2.8in]{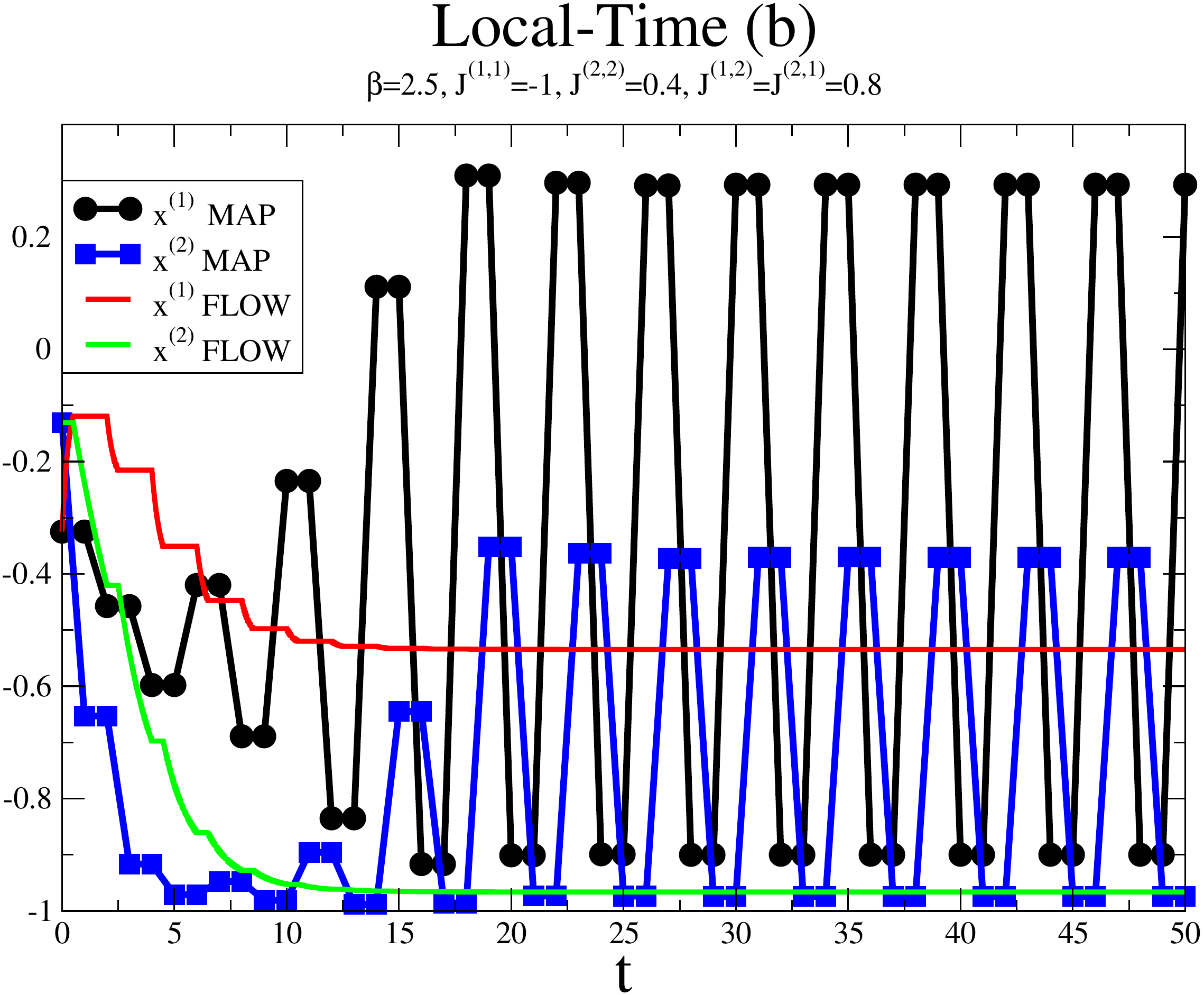}}\\ \vspace{0.5cm}
{\includegraphics[height=2.8in]{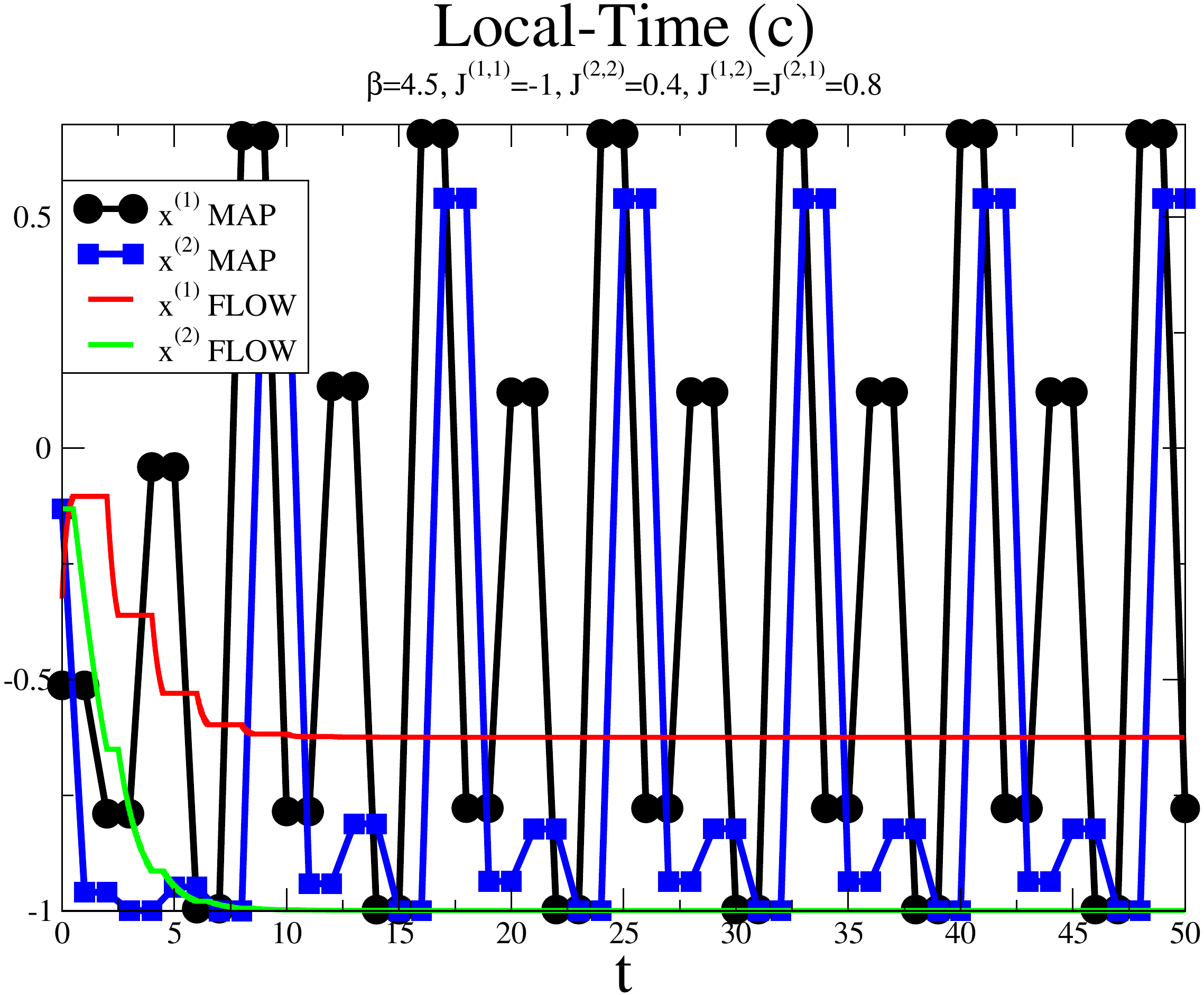}}
\captionof{figure}
{(Color online) Plot of $x^{(1)}(t)$ and $x^{(2)}(t)$, solution of the local-time flow, Eqs. (\ref{FFL1})-(\ref{FFL2}), 
and the local-time map, Eqs. (\ref{MML1})-(\ref{MML2}) (here with $\alpha=1$),  
$\beta=1.5$ (a), $\beta=2.5$ (b), and $\beta=4.5$ (c).
Initial conditions are the same in all cases. 
The lines of the map are guides for the eyes. 
  \label{fig3}
    }
\end{center}
\end{minipage}

\begin{figure}[htb]
\begin{center}
{\includegraphics[height=2.8in]{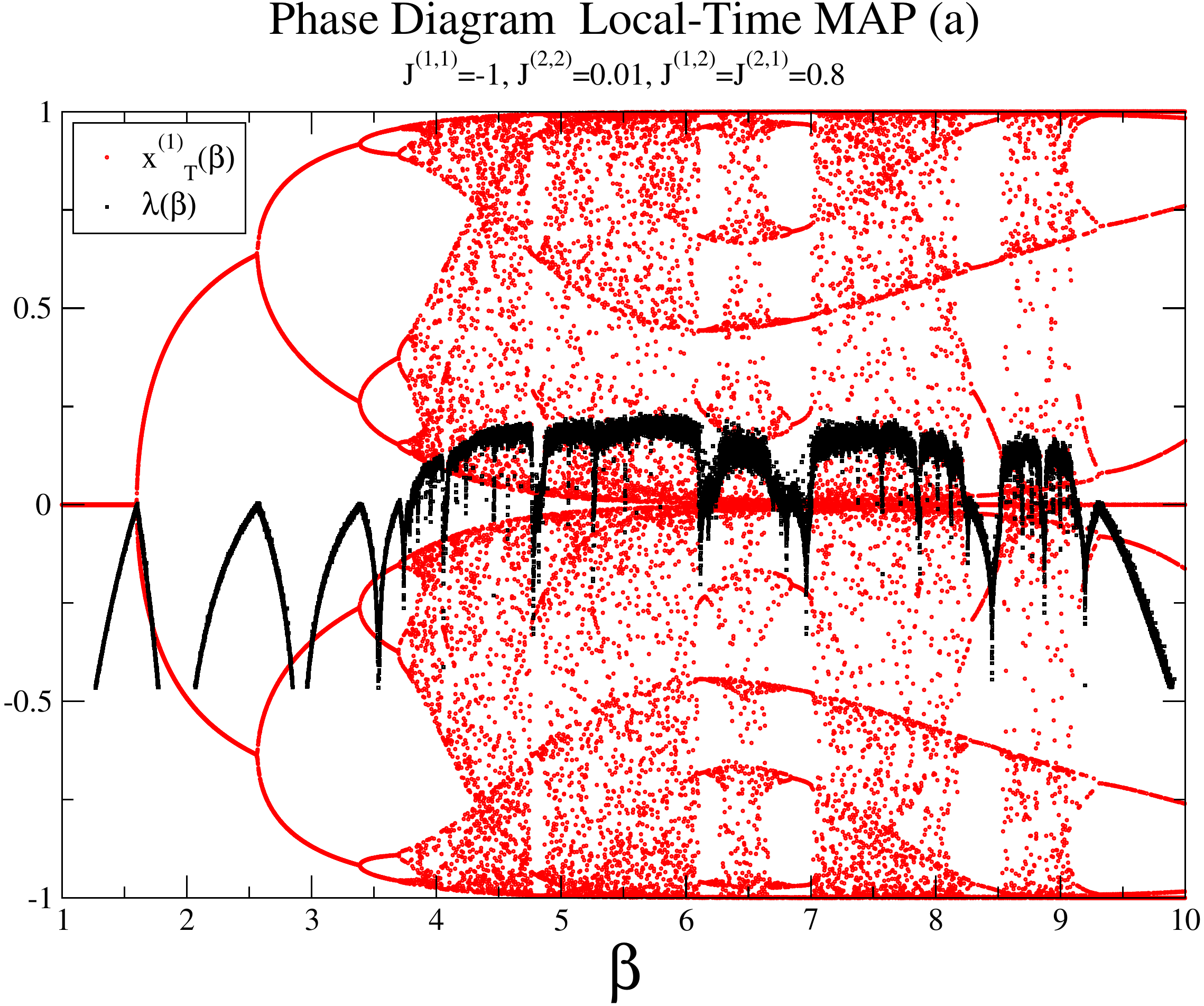}}\\ \vspace{0.5cm}
{\includegraphics[height=2.8in]{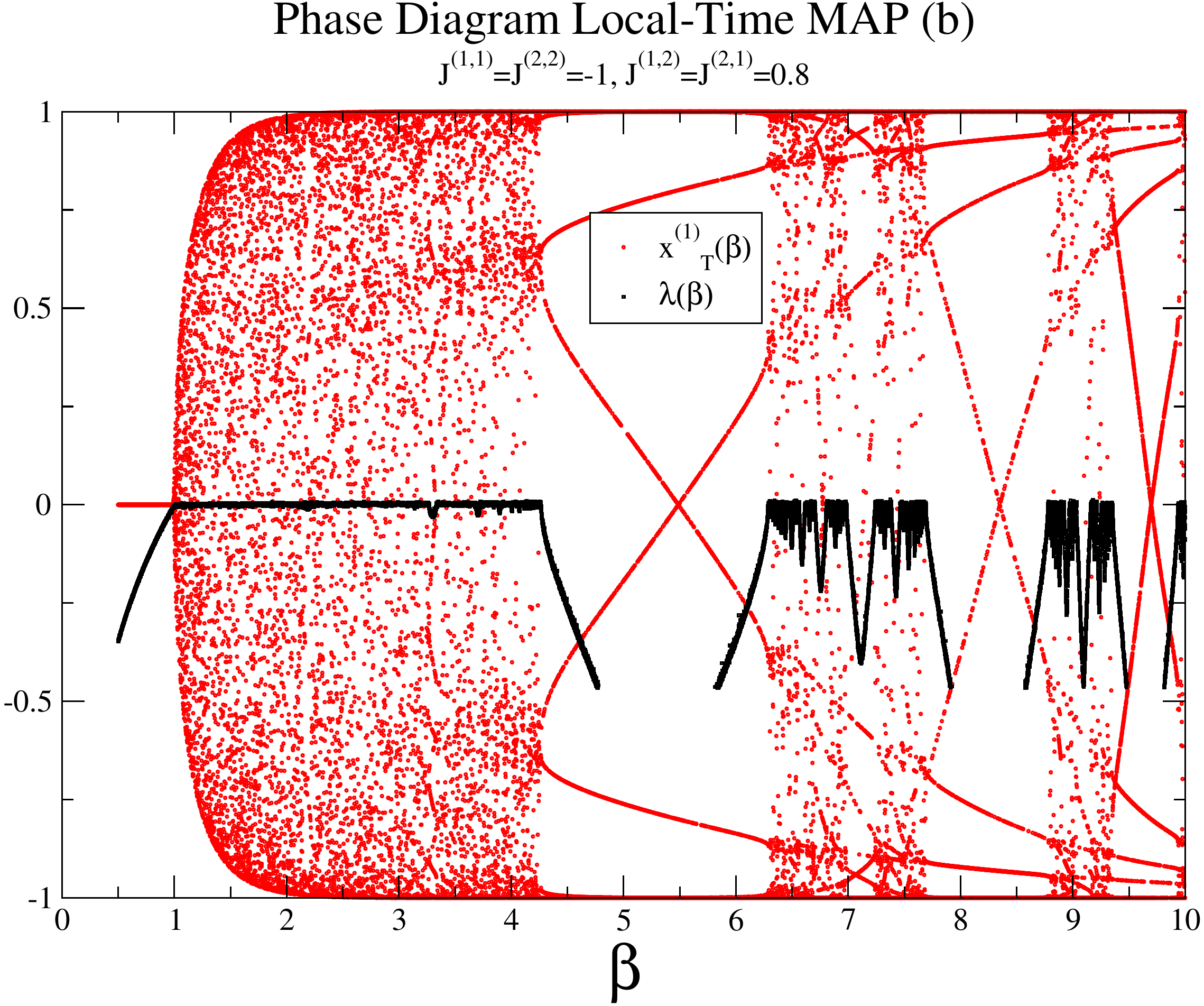}}
    \caption{(Color online) Plot of $x_T^{(1)}(\beta)$ (red dots), solution of the local-time map, Eqs. (\ref{MML1})-(\ref{MML2}) 
(here with $\alpha=1$) as a function of $\beta$, with $T$ high enough to remove
      temporary transients. In this case $T=10^4$. We plot also the maximum Lyapunov exponent $\lambda(\beta)$ 
(black highly irregular line) calculated from Eqs. (\ref{GlauberM2})-(\ref{Q}). 
      Each value of $\beta$ is in correspondence with a random initial condition for a total of $10^5$ samples.
Panel (a) is a case with bifurcation cascade, Panel (b) is a case with marginal chaos.  
  \label{fig4}
    }
  \end{center}
\end{figure}

\section{Chaos}
The case analyzed in Figs. (3) shows that, for large values of $\beta$, the local-time map can lead to erratic and, possibly, chaotic trajectories. 
In general, for each dynamics, the major quantity of interest is the Jacobian $\mathcal{J}^{(l,m)}_{t}$, where $l,m\in\{1,2\}$:
\subsection{FLOW}
\begin{eqnarray}
\label{GlauberF}
&& \mathcal{J}^{(l,m)}_{t}=
\left[1-\left(x^{(l)}_{t}\right)^2\right]\beta J^{(l,m)},
\end{eqnarray}
\subsection{MAP}
\begin{eqnarray}
\label{GlauberM}
&& \mathcal{J}^{(l,m)}_{t+1}=
\left[1-\left(x^{(l)}_{t}\right)^2\right]\beta J^{(l,m)},
\end{eqnarray}
\subsection{LOCAL-TIME FLOW}
\begin{eqnarray}
\label{GlauberF2}
&& \mathcal{J}^{(l,m)}_{t}=\left\{
\begin{array}{l}
\delta_{l,m}, \quad \quad \quad \mod[t+l-1,2]\neq 0, \\
\\
\left[1-\left(x^{(l)}_{t}\right)^2\right]\beta J^{(l,m)},
\quad \mod[t+l-1,2]=0,
\end{array}
\right.
\end{eqnarray}
\subsection{LOCAL-TIME MAP}
\begin{eqnarray}
\label{GlauberM2}
&& \mathcal{J}^{(l,m)}_{t+1}=\left\{
\begin{array}{l}
\delta_{l,m}, \quad \quad \quad \mod[t+l-1,2]\neq 0, \\
\\
\left[1-\left(x^{(l)}_{t}\right)^2\right]\beta J^{(l,m)},
\quad \mod[t+l-1,2]=0.
\end{array}
\right.
\end{eqnarray}
The knowledge of $\mathcal{J}^{(l,m)}_{t}$ allows to calculate the maximum Lyapunov exponent $\lambda$.
It turns out that in the map case (local-time or not), due to the fact that - via recurrence - we have the explicit 
solution of the Eqs., the evaluation of $\lambda$ is easier if compared to the flow case.
Since the flow does not show any signs of chaos or erratic trajectories, we can limit the evaluation of the
Lyapunov exponent to the map, where we can use
\begin{eqnarray}
\label{Lyapunov}
\lambda=\lim_{t\to\infty}\frac{1}{2t}\log(\Lambda_t),
\end{eqnarray}
where $\Lambda_t$ is the largest eigenvalue of the following matrix
\begin{eqnarray}
\label{Q}
\bm{Q}_t=\bm{L}_t \bm{L}_t^\dag, \quad \mathrm{where} \quad
\bm{L}_t=\prod_{t'=0}^t\bm{\mathcal{J}}_{t'}.
\end{eqnarray}
Given a set of couplings, we let the system
to evolve toward high enough values of $t=T$ in order to remove
temporary transients and repeat the numerical experiment for several values of $\beta$, each 
$\beta$ being associated to a random initial condition.
Hence, we plot $(x^{(1)}(T),x^{(2)}(T))$ as functions of $\beta$, and we indicate
these functions as $(x_{T}^{(1)}(\beta),x_{T}^{(2)}(\beta))$. 
In our examples and for the range of $\beta$ considered, $T \geq 10^3$ turns out to be enough high. 
In fact, the variable $\beta$ plays the role of a time-scaling:
the higher $\beta$, the higher the necessary transient $T$. 
In general, the functions $(x_{T}^{(1)}(\beta),x_{T}^{(2)}(\beta))$ look multi-valued
functions due to the existence of bifurcation points. There exist two kinds of bifurcations: 
doubling period and phase transition.
However, it is clear that in either case, doubling period or phase transition, the presence of bifurcations
increases the chance to develop chaos. Therefore, for our aims here,
it is more interesting, as well as highly more efficient, to show the plots
that include all the bifurcation points. 
Figs. (4a)-(4b) show two different scenarios.
The case of Panel (a) is a ``classical'' bifurcation cascade \cite{Logistic}, where, after a threshold $\beta_{Chaos}$, bifurcation takes place
increasingly, up to windows of stability. In the case of Panel (b) we observe a different situation which we call ``marginal chaos''.
In this case, after a threshold $\beta_{Marginal}$, the dynamics is characterized by totally erratic trajectories with no period, up
to windows of stability. Strictly speaking, even out of the windows of stability,  the system is not chaotic, since 
the maximum Lyapunov exponent is $\lambda=0$. However, in such regions, the system turns out to be critical over continuous intervals;
a situation which is not typical in models characterized by discrete symmetries like the Ising model. 

\section{Conclusions}
We have analyzed and compared four kinds of dynamics of a simple two component Ising-like system:
flow, map, local-time flow, and local-time map. Remarkably, even if only bilinear interactions are present,
the local-time map gives rise to erratic trajectories and, depending on the set of couplings, 
two chaotic scenarios take place: bifurcation cascades or ``marginal chaos'', 
\textit{i.e.}, criticality extended over continuous intervals of $\beta$, the time-scale parameter.
The analogous local-time flow does not present any of such behaviors, not even oscillations,
confirming all the scenarios we have already discussed in \cite{ChaosMO}.
We stress that this local-time update is not just a time-delay: here we have also a sleeping time
during which, in the case of some unfriendly couplings, there is an accumulation of frustration
among the system components, giving rise to instability and chaos, even for a system characterized
by  only bilinear interactions. As we have discussed in  \cite{ChaosMO}, we believe that maps with local-time updates
are a quite common feature in complex systems. Urgent issues will be to understand how this scenario
generalizes to large systems with many, say $q$ components, how the probability to have chaos changes
with $q$, and if the mean-field (or more precisely fully connected) picture remains robust.

\subsection{Acknowledgments}
Work supported 
by CNPq Grant PDS 150934/2013-0.
W. F. also acknowledges the Brazilian agencies CNPq and CAPES.

\end{document}